\documentclass[11pt,nofootinbib,showpacs,notitlepage]{revtex4-1}
\usepackage[T1]{fontenc}
\usepackage{babel}
\usepackage[utf8]{inputenc}
\usepackage{babel,amsmath,amssymb,mathrsfs,amsfonts,calligra,euscript,textcomp,graphicx,wrapfig,fontenc,pst-blur,tabu,lipsum,%
float,empheq,pst-node,environ,framed,ccaption,capt-of,enumitem,subcaption,enumitem,framed,txfonts,lmodern,cancel,diagbox,multirow,float,tabularx}
\usepackage[section]{placeins}
\definecolor{darkgreen}{rgb}{0,.7,0}
\usepackage[colorlinks,linkcolor=blue,citecolor=green,pagebackref=true]{hyperref}
\def\[{\left[}
\def\]{\right]}
\def\({\left(}
\def\){\right)}

\newcommand{\const}{\mathop{\rm const}\nolimits}

\newcommand{\m}{\mathop{\mathcal{M}}\nolimits}

\newcommand{\eq}[1]{\begin{equation}#1\end{equation}}

\begin{document}

\title{Stability analysis of compactification in 3-d order Lovelock gravity}

\author{Dmitry Chirkov}
\affiliation{Sternberg Astronomical Institute, Moscow State University, Moscow, Russia}

\author{Alexey Toporensky}
\affiliation{Sternberg Astronomical Institute, Moscow State University, Moscow, Russia}
\affiliation{Kazan Federal University, Kremlevskaya 18, Kazan 420008, Russia}
\begin{abstract}
    It is known that spatial curvature can stabilize extra dimensions in Lovelock gravity. In the present paper we study stability of the stabilization solutions in 3-d order Lovelock gravity.
    We show that in the case of negative spatial curvature of extra dimension space the stabilization solution is always stable. On the contrary, for positive spatial curvature the stability depends on the coupling constant values.
\end{abstract}
\maketitle

\section{Introduction}
The presence of higher-order curvature terms in the Lagrangian from one side was inspired by string theories. In 1974 Scherk and Schwarz~\cite{sch-sch} showed that the Lagrangian of the Virasoro-Shapiro model~\cite{VSh1, VSh2} contains $R^2$ and $R_{\alpha\beta}R^{\alpha\beta}$ terms. Later a presence of curvature-squared term of the form $R^{\alpha\beta\gamma\delta}R_{\alpha\beta\gamma\delta}$ was demonstrated~\cite{Candelas_etal} for the low-energy limit of the heterotic superstring theory~\cite{Gross_etal}. Next Zwiebach~\cite{Zwie} proved that the only combination of quadratic terms that gives a ghost-free solutions is the Gauss-Bonnet term $R^{\alpha\beta\gamma\delta}R_{\alpha\beta\gamma\delta}-4R_{\alpha\beta}R^{\alpha\beta}+R^2$. Zumino~\cite{zumino} extended Zwiebach's result on higher order curvature terms and put forward the idea that the low-energy limit Lagrangian of the unified theory should include a sum of different powers of curvature.

On the other hand, in 1970 Lovelock~\cite{Lovelock} discovered that as the number of space-time dimensions increases, in every new odd dimension it is possible to add a specific higher curvature power term to the classic Einstein-Hilbert Lagrangian such that the correction to the equations of motion will contain only derivatives  of the second order  of the metric coefficients. This make Lovelock's model a natural generalization of General Relativity (GR) to higher dimensions.

The simplest example of a large family of gravity theories known as Lovelock Gravities is Einstein-Gauss-Bonnet Gravity (EGB). Recent studies show that the dynamics of an anisotropic Universe in Einstein-Gauss-Bonnet gravity can be much reacher than it is possible in GR. In particular, the presence of cosmological constant $\Lambda$ does not ultimately lead to isotropic multidimensional de Sitter solution. Another non-singular possibility is an anisotropic solution where Hubble parameters $H_i$ are constant, but can be different for different $i$. On the other hand, all $H_i$ can not be completely different: it was shown that such solutions  exist only if the space has isotropic subspaces (three at most), so that the set of $H_i$ is divided into two or three groups with equal $H_i$ within a group~\cite{ChPavTop-ModPhysLet-2014,I,ChPavTop-GenRelGrav-2015}. It is noteworthy that this splitting is not done by the hand of an investigator, but appears in a natural way from equations of motion as a condition for such solutions to exist.

Explicit forms of the solutions in question show that the most typical situation (apart from de Sitter solution) is splitting of space into two isotropic subspaces with different Hubble parameters which we denote here as $H$ and $h$. Stability analysis indicates that for such solution to be stable the overall volume of space should increase~\cite{Iva-EurPhys-2016}. This leads to a condition $dh+(n-d)H>0$ for $n$-dimensional space splitted into $d$- and $n-d$-dimensional isotropic subspaces. Apart from this condition there is another one, which is shown to be violated if $d=1$~\cite{ChTop-GravCosmol-2017}, other possibilities are allowed except for some special combinations of coupling constants. In~\cite{split} it was shown numerically that such a decomposition is indeed a typical outcome of an anisotropic cosmological evolution if a non-zero cosmological constant is present in the action, with other possibilities being a non-standard singularity or an oscillatory regime.

So that the outlined results justify to consider a metric splitted into a product of two isotropic subspaces. The most interesting case is with $H>0$ in 3-dimensional subspace and $h<0$ in the $n-3$ dimensional subspace since such a combination would be useful for a compactification scenario describing "our" big 3-dimensional world and $n-3$-dimensional "inner" subspace. What is, however, still needed is a stabilization of the "inner" dimensions. In a set of papers~\cite{CanGiaPav-PhysRev-2013,CanGiaPav-GenRelGrav-2014,CanGiaPav-GravCosmol-2018} it was shown that such a stabilization can be achieved by introducing a negative spatial curvature for the "inner" space. As for the large subspace, possible spatial curvature does not change much since this subspace is expanding and the dynamical role of the spatial curvature decreases. So, the initially anisotropic Universes naturally evolves into a product of large isotropic "our" sub-Universe and a stabilized isotropic "inner" space -- in what follows we refer to such solutions as a compactification solution. Since $H>0$ for the large subspace, an inside observer would indicate the presence of an effective cosmological constant $\Lambda_{eff}$ in "our" world. Choosing the coupling constant of the theory satisfying an additional relation it is possible to set this $\Lambda_{eff}$ to arbitrary small value and get a standard Friedmann equation for evolution of the large subspace~\cite{CanGiaTroWil-PhysRev-2009}.

Stability of compactified solutions have been considered in~\cite{Pav-PhysRev-2015,Iva-EurPhys-2016,Iva-GravCosmol-2016,ErnIva-EurPhys-2017,ChTop-GravCosmol-2017} for both negative and positive curvature of the "inner" space. It was found that for negative curvature the solution with constant volume of the inner space is stable when it exists, however, for positive curvature the stability requires rather tough restrictions of possible coupling constants, and these restrictions are the more severe the bigger number of extra dimensions is considered. What is even worse, the case of zero effective cosmological constant in large dimensions is always unstable~\cite{ChPav-MosPhysLet-2021}. In Lovelock gravity, starting from $n=6$ the 3-d term is added to the Gauss-Bonnet term. In~\cite{ChGiaTop-GenRelGrav-2018} the stabilization of extra dimensions have been checked numerically for negative spatial curvature in the presence of 3-d Lovelock term. The goal of the present paper is to study stability of stabilization solutions with the cubic term for both possible cases of spatial curvature of extra dimensions.

The structure of the paper is the following: section~\ref{EoM} presents the action, the Lagrangian and the equations of motion; section~\ref{num-analysis} is devoted to the numerical analysis of stability and in the last section the conclusions will be given.

\section{Action and equations of motion\label{EoM}}
We consider $(D+4)$-dimensional spacetime $\mathcal{M}=\m_4\times \m_D$ where $\m_4$ is a flat Friedman-Robertson-Walker manifold with scale factor $a(t)$, $\m_D$ is a $D$-dimensional Euclidean compact constant curvature manifold with scale factor $b(t)$ and curvature $\gamma_{D}$.

Lovelock action under consideration reads
\eq{S=\int_{\mathcal{M}}d^{D+4}x\sqrt{|g|}\left\{R+\alpha L_{(2)}+\beta L_{(3)}-2\Lambda\right\},\label{action}}
where $|g|$ is the determinant of metric tensor; $\Lambda$ is the cosmological term; $\alpha$ and $\beta$ are the coupling constants; $L_{(2)}$ is quadratic Lovelock term\footnote{Hereafter Greek indices run from 0 to D, while Latin one from
1 to D unless otherwise stated}:
\eq{L_{(2)}=R^2-4R_{\beta}^{\phantom{\beta}\alpha}R_{\alpha}^{\phantom{\alpha}\beta}
+R_{\gamma\delta}^{\phantom{\gamma\delta}\alpha\beta}R_{\alpha\beta}^{\phantom{\alpha\beta}\gamma\delta}}
and $L_{(3)}$ is cubic Lovelock term:
\eq{\begin{split}
L_{(3)}=&-R^3+12RR_{\beta}^{\phantom{\beta}\alpha}R_{\alpha}^{\phantom{\alpha}\beta}
-3RR_{\gamma\delta}^{\phantom{\gamma\delta}\alpha\beta}R_{\alpha\beta}^{\phantom{\alpha\beta}\gamma\delta}
-16R_{\beta}^{\phantom{\beta}\alpha}R_{\gamma}^{\phantom{\gamma}\beta}R_{\alpha}^{\phantom{\alpha}\gamma}
+24R_{\gamma}^{\phantom{\gamma}\alpha}R_{\delta}^{\phantom{\delta}\beta}R_{\alpha\beta}^{\phantom{\alpha\beta}\gamma\delta}+\\
        &+24R_{\beta}^{\phantom{\beta}\alpha}R_{\delta\varepsilon}^{\phantom{\delta\varepsilon}\beta\gamma}R_{\alpha\gamma}^{\phantom{\alpha\gamma}\delta\varepsilon}
         +2R_{\gamma\delta}^{\phantom{\gamma\delta}\alpha\beta}R_{\varepsilon\zeta}^{\phantom{\varepsilon\zeta}\gamma\delta}
R_{\alpha\beta}^{\phantom{\alpha\beta}\varepsilon\zeta}
-8R_{\gamma\varepsilon}^{\phantom{\gamma\varepsilon}\alpha\beta}R_{\alpha\zeta}^{\phantom{\alpha\zeta}\gamma\delta}
R_{\beta\delta}^{\phantom{\beta\delta}\varepsilon\zeta}
    \end{split}}
$R,R_{\beta}^{\phantom{\beta}\alpha},R_{\alpha\beta}^{\phantom{\alpha\beta}\gamma\delta}$ are the $(D+4)$-dimensional scalar curvature, Ricci tensor and Riemann tensor respectively.

We choose the ansatz for the metric as follows
\eq{ds^2=-dt^2+a(t)^2d\Sigma^2_{3}+b(t)^2d\Sigma^2_{D}\label{metric}}
where $d\Sigma^2_{3}$ stand for the metric of 3-dimensional plane, $d\Sigma^2_{D}$ stand for the metrics of $D$-dimensional manifold with constant curvature.

Substituting metric into the action~(\ref{action}), varying it with respect to $a(t)$ and $b(t)$ and introducing the Hubble parameter $H\equiv\frac{\dot{a}}{a}$, we obtain constraint~(\ref{constraint}) and equations of motion~(\ref{Eq1})-(\ref{Eq2}):
\eq{\begin{split}
       &\frac{3}{D+1}\left(\frac{H b'(D+1)!}{b (D-1)!}+\frac{H^2(D+1)!}{D!}+\frac{(\gamma_{D}+b'^2)(D+1)!}{6b^2(D-2)!}\right)+ \\
         & +3D\alpha\Biggl(\frac{(\gamma_{D}+b'^2)^2(D-1)!}{6b^4(D-4)!}+\frac{2H^2(\gamma_{D}+b'^2)(D-1)!}{b^2(D-2)!}+
      \frac{4H^3b'}{b}+ \\
         & \hspace{4.8cm}+\frac{4H^2b'^2(D-1)!}{b^2(D-2)!}+\frac{2Hb'(\gamma_{D}+b'^2)(D-1)!}{b^3(D-3)!}\Biggr)+ \\
         & +3D(D-1)(D-2)\beta\Biggl(\frac{(\gamma_{D}+b'^2)^3(D-3)!}{6b^6(D-6)!}+\frac{3H^2(\gamma_{D}+b'^2)^2(D-3)!}{b^4(D-4)!}+
      \frac{3Hb'(\gamma_{D}+b'^2)^2(D-3)!}{b^5(D-5)!}+ \\
         & \hspace{4cm}+\frac{8H^3b'^3}{b^3}+\frac{12H^2b'^2(\gamma_{D}+b'^2)(D-3)!}{b^4(D-4)!}+\frac{12H^3b'(\gamma_{D}+b'^2)}{b^3}\Biggr)=\Lambda
    \end{split}
\label{constraint}}

\eq{\begin{split}
       &\frac{1}{D+1}\left(\frac{2H b'(D+1)!}{b (D-1)!}+\frac{H^2(D+1)!}{D!}+\frac{(\gamma_{D}+b'^2)(D+1)!}{2b^2(D-2)!}+\frac{b''(D+1)!}{b(D-1)!}+\frac{2(H'+H^2)(D+1)!}{D!}\right)+ \\
         &+D\alpha\Biggl(\frac{(\gamma_{D}+b'^2)^2(D-1)!}{2b^4(D-4)!}+
      \frac{8b''b'H(D-1)!}{b^2(D-2)!}+\frac{4(\gamma_{D}+b'^2)(H'+H^2)(D-1)!}{b^2(D-2)!}+\\
         &\hspace{1.3cm} +\frac{4H^2b''}{b}+\frac{4Hb'(\gamma_{D}+b'^2)(D-1)!}{b^3(D-3)!}+\frac{4H^2b'^2(D-1)!}{b^2(D-2)!}+\\
         &\hspace{1.3cm} +\frac{2H^2(\gamma_{D}+b'^2)(D-1)!}{b^2(D-2)!}+\frac{8(H'+H^2)Hb'}{b}+\frac{2b''(\gamma_{D}+b'^2)(D-1)!}{b^3(D-3)!}\Biggr)+ \\
         & +D(D-1)(D-2)\beta\Biggl(\frac{(\gamma_{D}+b'^2)^3(D-3)!}{2b^6(D-6)!}+\frac{3H^2(\gamma_{D}+b'^2)^2(D-3)!}{b^4(D-4)!}+
      \frac{6Hb'(\gamma_{D}+b'^2)^2(D-3)!}{b^5(D-5)!}+ \\
         &\hspace{1.1cm}+\frac{24b''H^2b'^2}{b^3}+\frac{12H^2b'^2(\gamma_{D}+b'^2)(D-3)!}{b^4(D-4)!}+\frac{24(H'+H^2)Hb'(\gamma_{D}+b'^2)}{b^3}+\frac{12b''H^2(\gamma_{D}+b'^2)}{b^3}+ \\ &\hspace{1.1cm}+\frac{3b''(\gamma_{D}+b'^2)^2(D-3)!}{b^5(D-5)!}+\frac{6(H'+H^2)(\gamma_{D}+b'^2)^2(D-3)!}{b^4(D-4)!}+
         \frac{24b''b'H(\gamma_{D}+b'^2)(D-3)!}{b^4(D-4)!}\Biggr)=\Lambda
    \end{split}
\label{Eq1}}

\eq{\begin{split}
       &\frac{3}{D+1}\left(\frac{H b'(D+1)!}{b (D-2)!}+\frac{H^2(D+1)!}{(D-1)!}+\frac{(\gamma_{D}+b'^2)(D+1)!}{6b^2(D-3)!}+\frac{b''(D+1)!}{3b(D-2)!}+\frac{(H'+H^2)(D+1)!}{(D-1)!}\right)+ \\
         &+3D\alpha\Biggl(\frac{(\gamma_{D}+b'^2)^2(D-1)!}{6b^4(D-5)!}+\frac{2H^2(\gamma_{D}+b'^2)(D-1)!}{b^2(D-3)!}+
      \frac{4b''b'H(D-1)!}{b^2(D-3)!}+\frac{4H^3b'(D-1)!}{b(D-2)!}+\\
         & \hspace{1.1cm}\frac{4H^2b''(D-1)!}{b(D-2)!}+\frac{2Hb'(\gamma_{D}+b'^2)(D-1)!}{b^3(D-4)!}+\frac{4H^2b'^2(D-1)!}{b^2(D-3)!}+
         \frac{8(H'+H^2)Hb'(D-1)!}{b(D-2)!}+ \\
         &\hspace{1.1cm}\frac{2b''(\gamma_{D}+b'^2)(D-1)!}{3b^3(D-4)!}+\frac{2(H'+H^2)(\gamma_{D}+b'^2)(D-1)!}{b^2(D-3)!}+4H^2(H'+H^2)\Biggr)\\
         & +3D(D-1)(D-2)\beta\Biggl(\frac{(\gamma_{D}+b'^2)^3(D-3)!}{6b^6(D-7)!}+\frac{3H^2(\gamma_{D}+b'^2)^2(D-3)!}{b^4(D-5)!}+
              \frac{3Hb'(\gamma_{D}+b'^2)^2(D-3)!}{b^5(D-6)!}+ \\
         &\hspace{1.1cm}\frac{12H^2b'^2(\gamma_{D}+b'^2)(D-3)!}{b^4(D-5)!}+\frac{24(H'+H^2)Hb'(\gamma_{D}+b'^2)(D-3)!}{b^3(D-4)!}+
         \frac{24b''H^2b'^2(D-3)!}{b^3(D-4)!}+ \\
         & \hspace{1.1cm}\frac{12b''H^2(\gamma_{D}+b'^2)(D-3)!}{b^3(D-4)!}+\frac{12H^2(H'+H^2)(\gamma_{D}+b'^2)}{b^2}+\frac{24H^2b'^2(H'+H^2)}{b^2}+\\
         &\hspace{1.1cm}\frac{12H^3b'(\gamma_{D}+b'^2)(D-3)!}{b^3(D-4)!}+\frac{8H^3b'^3(D-3)!}{b^3(D-4)!}+\frac{24b''b'H^3}{b^2}\\
         & \hspace{1.1cm}\frac{b''(\gamma_{D}+b'^2)^2(D-3)!}{b^5(D-6)!}+\frac{3(H'+H^2)(\gamma_{D}+b'^2)^2(D-3)!}{b^4(D-5)!}+
         \frac{12b''b'H(\gamma_{D}+b'^2)(D-3)!}{b^4(D-5)!}\Biggr)=D\Lambda
    \end{split}
\label{Eq2}}

\section{Stability analysis for the case $D=7$\label{num-analysis}}
Henceforward we assume that $D=7$. Substituting $D=7$ into the equations (\ref{Eq1})-(\ref{Eq2}) and solving it with respect to higher derivatives we obtain autonomous system of ordinary differential equations
\eq{
\left\{
\begin{array}{l}
\dot{b}=u \\
\dot{u}=F_1(b,u,H) \\
\dot{H}=F_2(b,u,H)
\end{array}
\right.
\label{system-buH}}
We do not write these equations down in an explicit way because of their cumbersomeness.

Compactification scenario suggests that at late times 3 dimensions describing "our real world" are expanding at an accelerating rate whereas the extra dimensions tend to a constant size. It means that $H'(t),b'(t),b''(t)\rightarrow0$; $b(t)\underset{t\rightarrow\infty}{\longrightarrow}\nolinebreak b_0$, $H(t)\underset{t\rightarrow\infty}{\longrightarrow}\nolinebreak H_0$, where $b_0,H_0=\const$. Since the value of the observed cosmological constant $H\sim10^{-132}$ in fundamental units, physically realistic regime implies that $H_0\approx0$.

Substituting $b''=b'=H'=H=0,\;b=b_0,\;D=7$ into constraint~(\ref{constraint}) and equations of motion~(\ref{Eq1})-(\ref{Eq2}), we get equations which we call \emph{asymptotic equations} in what follows:
\eq{\Lambda\,b_0^{6}-42\,\gamma_D\,b_0^{4}-840\,\alpha\,\gamma_D^{2}b_0^{2}-5040\,\beta\,\gamma_D^{3}=0}
\eq{\Lambda\,b_0^{6}-30\,\gamma_D\,b_0^{4}-360\,\alpha\,\gamma_D^{2}b_0^{2}-720\,\beta\,\gamma_D^{3}=0}
One of the asymptotic equations of motion coincides with the constraint, so we have only two independent equations. Assuming $b_0>0$ we solve these equation with respect to $\Lambda$ and $b_0$:
\eq{\Lambda={\frac {30\,\gamma_D\,({{b_0}}^{4}+12\,\alpha\,\gamma_D\,{{b_0}}^{2}+24\,\beta\gamma_D^{2})}{{{b_0}}^{6}}}\label{Lambda}}

\begin{table}[!h]
\begin{center}
\caption{Conditions of existing of $b_0$}
\label{table.solutions.5+1}
  \begin{tabular}{|c|c|c|}
    \hline
    & $\gamma_D>0$ & $\gamma_D<0$ \\
    \hline
    $b_0=\sqrt{-20\alpha\gamma_D+2\gamma_D\sqrt{100\alpha^2-90\beta}}$ &
    $\begin{array}{c}
       \beta<0,\;\;\alpha\in\mathbb{R} \\
       0<\beta\leqslant\frac{10}{9}\alpha^2,\;\;\alpha<0
     \end{array}$ &
    $0<\beta\leqslant\frac{10}{9}\alpha^2,\;\;\alpha>0$ \\
    \hline
    $b_0=\sqrt{-20\alpha\gamma_D-2\gamma_D\sqrt{100\alpha^2-90\beta}}$ &
    $0<\beta\leqslant\frac{10}{9}\alpha^2,\;\;\alpha<0$ &
    $\begin{array}{c}
       \beta<0,\;\;\alpha\in\mathbb{R} \\
       0<\beta\leqslant\frac{10}{9}\alpha^2,\;\;\alpha>0
     \end{array}$ \\
    \hline
  \end{tabular}
\end{center}
\end{table}

%\eq{\begin{array}{c}
%      \;\;\mbox{for}\;\;\beta<0,\;\;\alpha\in\mathbb{R},\;\;\gamma_D>0 \vspace{0.2cm}\\
%      \;\;\mbox{for}\;\;\beta<0,\;\;\alpha\in\mathbb{R},\;\;\gamma_D<0 \vspace{0.2cm}\\
%      \left[
%      \begin{array}{c}
%        b_0=\sqrt{-20\alpha\gamma_D+2\gamma_D\sqrt{100\alpha^2-90\beta}} \\
%        b_0=\sqrt{-20\alpha\gamma_D-2\gamma_D\sqrt{100\alpha^2-90\beta}}
%      \end{array}\right.
%      \;\;\mbox{for}\;\;0<\beta\leqslant\frac{10}{9}\alpha^2,\;\;\alpha<0
%    \end{array}
%\label{b0}}

Compactified solution $\bigl\{b(t)\equiv b_0,\; u(t)\equiv 0,\; H(t)\equiv 0\bigr\}$ is a fixed point of the system~(\ref{system-buH}). Stability of a fixed point of a system of ODEs is determined by the sign of real part of eigenvalues of the Jacobian matrix evaluated at the this point; a fixed point is asymptotically stable if all eigenvalues have negative real parts.

\begin{figure}[!t]
\begin{minipage}[h]{.32\linewidth}
\center{\includegraphics[width=\linewidth]{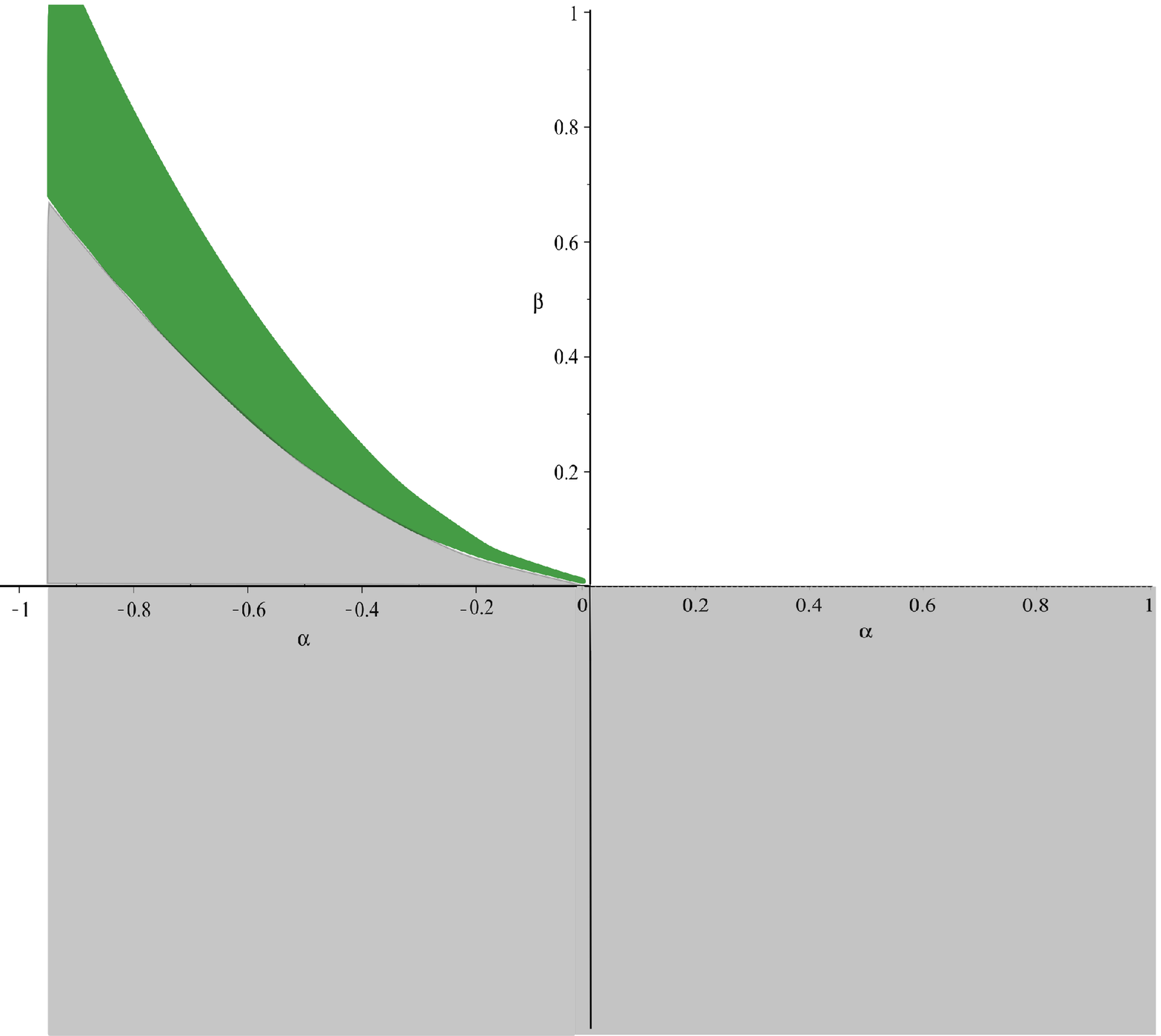} \\ a)}
\end{minipage}
\begin{minipage}[h]{.32\linewidth}
\center{\includegraphics[width=\linewidth]{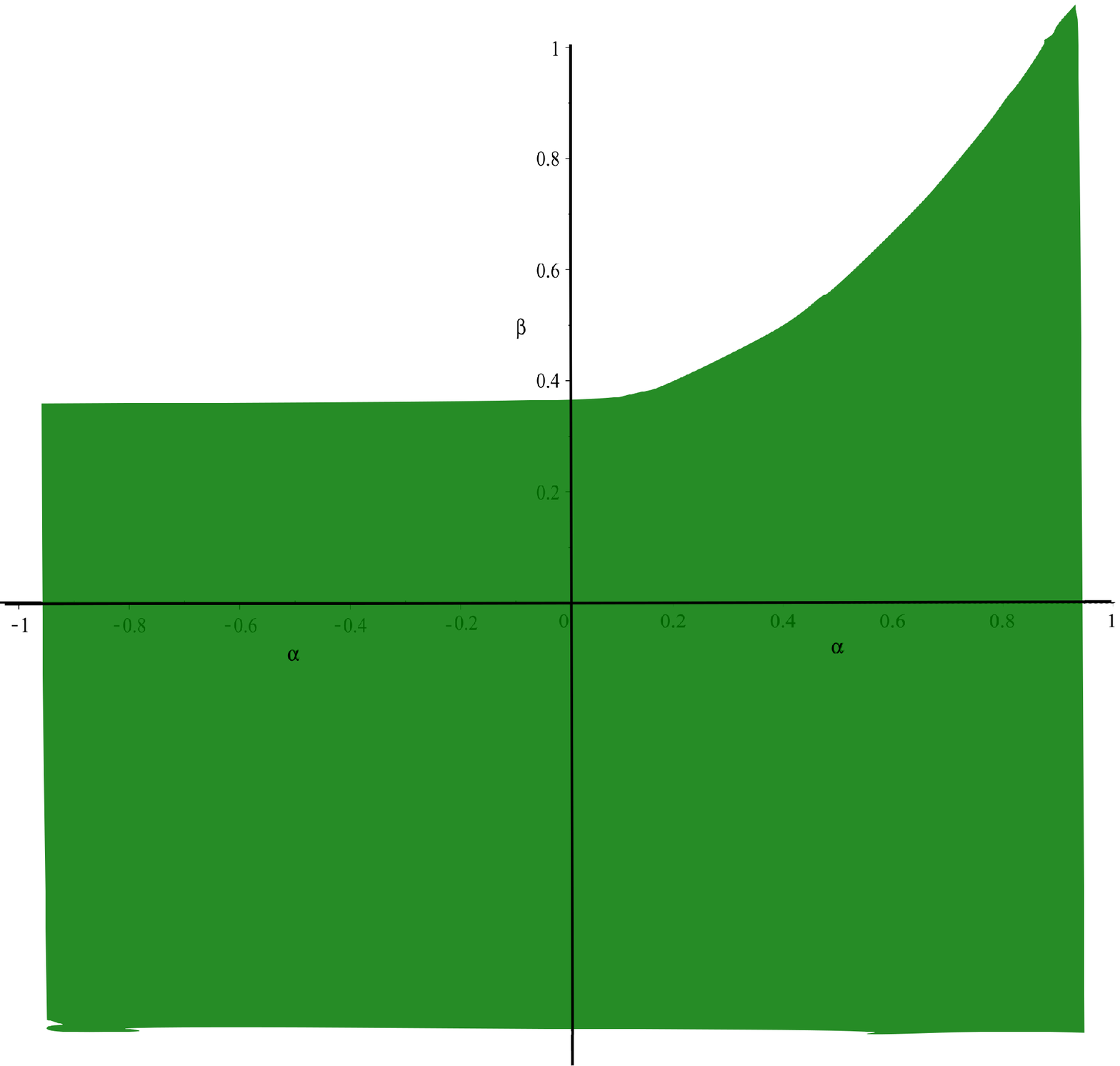} \\ b)}
\end{minipage}
\begin{minipage}[h]{.32\linewidth}
\center{\includegraphics[width=\linewidth]{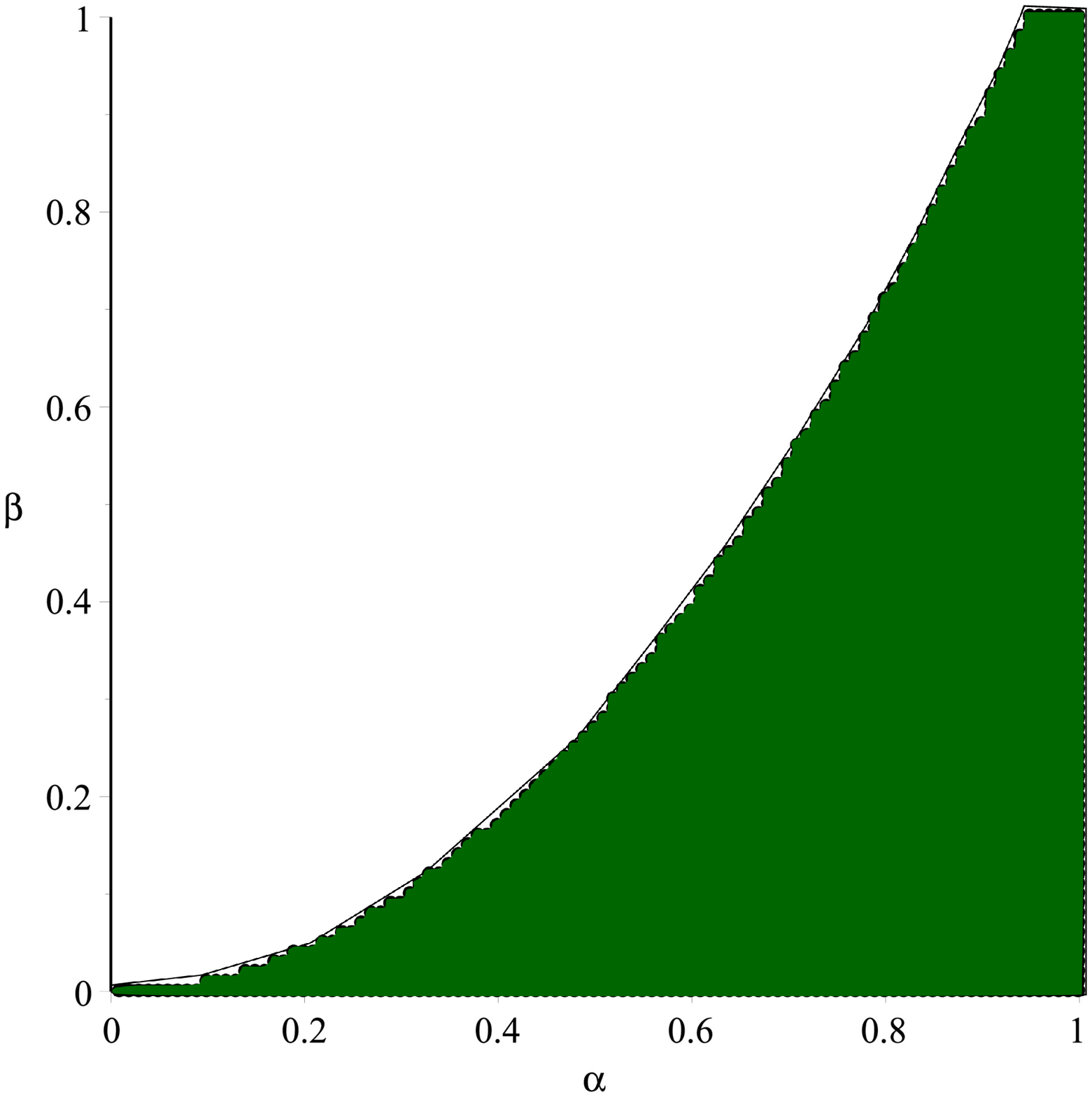} \\ c)}
\end{minipage}
\caption{\footnotesize The distribution of stable compactified solutions over the coupling constants $\alpha$ and $\beta$ for $H_0=10^{-15}$ for  a) positive curvature; b) negative curvature, the branch $b_0=\sqrt{20\alpha+2\sqrt{100\alpha^2-90\beta}}$; c) negative curvature, the branch $b_0=\sqrt{20\alpha-2\sqrt{100\alpha^2-90\beta}}$. Green areas: stable solutions; grey area: unstable solutions; white area: solutions does not exist.}
\label{stable-compactification}
\end{figure}

Analytical expressions for the corresponding eigenvalues can be written down, though they are too  combersom for using them in further studies. So that,
we check the stability condition by numerical evaluation. We also note that substitution $H=0$ lead to all zero eigenvalues or to one zero and two equal in absolute value but opposite in sign eigenvalues (the same is true for the case of negative curvature). In order to be safe from small computer errors, instead of exact $H=0$
we substitute $u=0,\; H\sim 10^{-16}$ as well as expressions for $\Lambda$ and $b_0$ into the Jacobian matrix of the system~(\ref{system-buH}), so that each element of this matrix becomes a function of the coupling constants $\alpha$ and $\beta$. After that we make a mesh with the coupling constants and evaluate eigenvalues for each pair $(\alpha,\beta)$.  It should also be noted that for positive curvature we see stable solutions for $b_0=\sqrt{-20\alpha+2\sqrt{100\alpha^2-90\beta}}$ with "plus" sign before the square root, and we do not see any stable solution for the branch of $b_0$ with "minus" sign before the square root ($b_0=\nolinebreak\sqrt{-20\alpha-2\sqrt{100\alpha^2-90\beta}}$) whereas for negative curvature we get stable solutions for both branches of $b_0$. Figure~\ref{stable-compactification} illustrates the distribution of stable compactified solutions over the coupling constants $\alpha$ and $\beta$ for $H_0=10^{-15}$ for positive and negative curvature. Green areas of the figures represent stable solutions; grey areas - unstable solutions; in white areas solutions does not exist at all. The result we see for negative curvature confirms the result obtained in our previous paper~\cite{ChGiaPavTop-EurPhys-2021}: all existing solutions are always stable.

One can see that for $\beta=0$ (Einstein-Gauss-Bonnet model) there are no stable compactified solutions with positive curvature, however they exist with negative curvature. These result are in agreement with the results obtained earlier (see~\cite{ChPav-MosPhysLet-2021} for details). For non-zero $\beta$ stable solutions exist for any sign of the spatial curvature.

\begin{figure}[!t]
\begin{minipage}[h]{.49\linewidth}
\center{\includegraphics[width=0.6\linewidth]{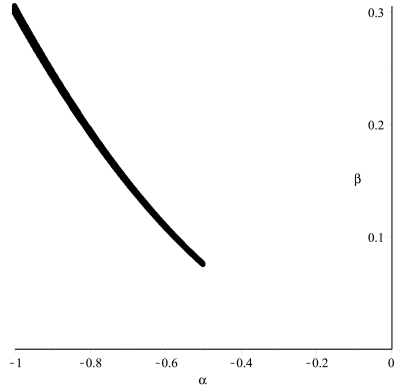}}
\end{minipage}
\caption{\footnotesize Region of stability for $D=13$ extra dimensions (positive curvature)}
\label{changing-region-positive-curvature}
\end{figure}

Note that as the number of extra dimensions increases the number of stable compactified solutions with positive curvature decreases (see Fig.~\ref{changing-region-positive-curvature}); in the case of negative curvature we do not observe this and  the regions of stability do not change. In the case of Einstein-Gauss-Bonnet model there exists a similar situation in the general case when we abandon the $H \sim 0$ condition and compactification for positive curvature becomes possible~\cite{ChGiaPavTop-EurPhys-2021} .

\section {Conclusions}

We have studied the stability of compactification scenario in 3-d order Lovelock gravity with curved extra dimension subspace.
We considered both negative and positive possibilities for the spatial curvature of the "inner" space. In the present paper
we restricted ourselves by the case when effective cosmological constant in "big" dimensions subspace is small.

Our results indicate that some restrictions known for Einstein-Gauss-Bonnet gravity (i.e. without the 3-d Lovelock term)
can be lifted, in particular the scenario with positive spatial curvature of "inner" space being impossible without
the 3-d term appears to be possible for non-zero 3-d term. However, if we plot coupling constants ranges needed for such scenario
to realize, they form rather narrow band, moreover, the width of such zone decreases with number of extra dimensions increasing.
This is similar to compactification conditions in  EGB gravity in the general case (when effective cosmological constant in
"big" dimensions can be large, as we mentioned zero cosmological constant is incompatible with positive spatial curvature
of extra dimensions space): compactification scenario with positive spatial curvature is possible, but requires some fine-tuning
of coupling constants and the bigger number of extra dimensions is, the more severe this fine-tuning becomes.

As for the compactification with negative spatial curvature of the "inner" dimension space, the results are qualitatively the
same as for EGB gravity: if a compactification solution exists, it is always stable; zone of coupling constants needed for
compactification solution to exist is large and does not depend on the number of extra dimensions.

We also note that when for a stable point we choose $\Lambda$ so that $H$ vanishes, all three eigenvalue also vanish.
This means that initial perturbation would not disappear and instead would oscillate near the exact solution. Such oscillations
have been found in EGB gravity with $H=0$, moreover, numerical integration have shown that in the presence of ordinary matter
the oscillations decays in time \cite{CanGiaPav-GravCosmol-2018}. We can hope that this is still valid for 3-d order
Lovelock gravity, though the case of $H=0$ exactly needs a special treatment and we leave it for a future work.

\section*{Acknowledgements}
The work of AT  have been supported by the RFBR grant 20-02-00411. Authors are grateful to Alex
Giacomini for discussions.

\end{document}